# Non linear system become linear system

**Prof. PhD Petre Bucur,**
**Head of Computer Science Department**
**Asoc. Prof. PhD Lucian Luca,**
**Dean**
**Faculty of Computers and Applied Computer Science**
**„Tibiscus" University of Timisoara, Romania**

**REZUMAT.** Lucrarea se referă la teoria şi practica sistemelor cu privire la sisteme neliniare şi aplicaţiile lor. Se are în vedere integrarea acestor sisteme pentru elaborarea răspunsului lor, precum şi evidenţierea unor particularităţi deosebite.

## 1 Nonlinear systems transforms in another system (linear)

Non linear systems described many process of nature or other area. One of them is formulated in normal Cauchy form, as

$$\begin{cases} \dfrac{dx}{dt} = -x - x \cdot y^2 + c_2 \cdot z + c_1 \\ \dfrac{dy}{dt} = \dfrac{x + x \cdot y^2 - y}{c_3} \\ \dfrac{dz}{dt} = \dfrac{y - z}{c_4} - \dfrac{x \cdot y \cdot z}{c4} \end{cases}$$

Initial conditions of this system are in form

$$t = t_0 = 0 \; ; \; x(t_0) = 0 \; ; \; y(t_0) = 0 \; ; \; z(t_0) = 0$$

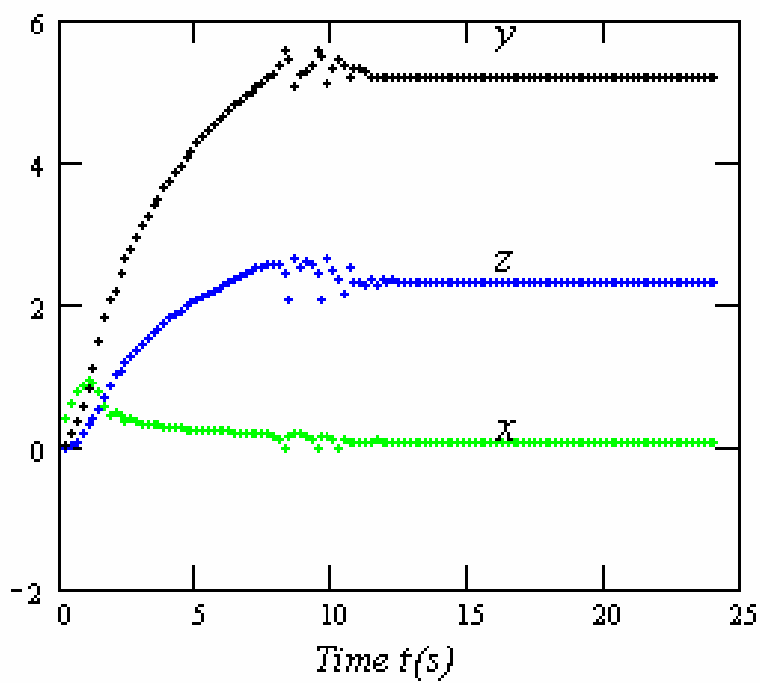

***Figure 1.1** nonlinear system pass to another system from beginning with initial conditions, bifurcations, chaos, bifurcations and finally linear system*

In the name of coefficients and conditions, after many search obtained the best transformation of a nonlinear system in linear system.

The phenomena that produced in time consists in the begin of dynamical system from initial conditions, after any time the system passed in bifurcations, chaos another bifurcations and recombined to a linear system. *Is fantastical!*

The system is integrated by numerical method Runge Kutta or Euler II for any coefficients' values. In this way, is obtained a variety of graphical forms. For example

$$c_{11} = 1 \; ; \; c_2 = 1{,}67515 \; ; \; c_3 = 1 \; ; \; c_4 = 1$$

With these coefficients is possible to obtain the same diagrams as in figure 1.1, but for a long time.

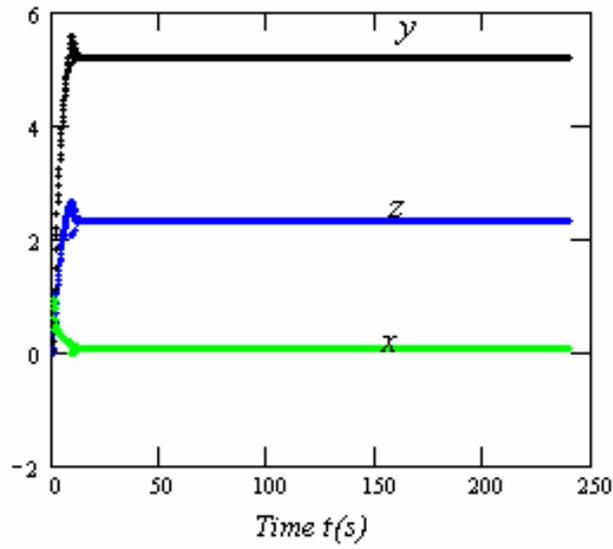

***Figure 1.2***  *illustrates the long time of the new system*

In other conditions and coefficients' values the initial nonlinear system has a bifurcation with

$$c_1 = 25 \; ; \; c_2 = 0 \; ; \; c_3 = 0,05017 \; ; \; c_4 = 1$$

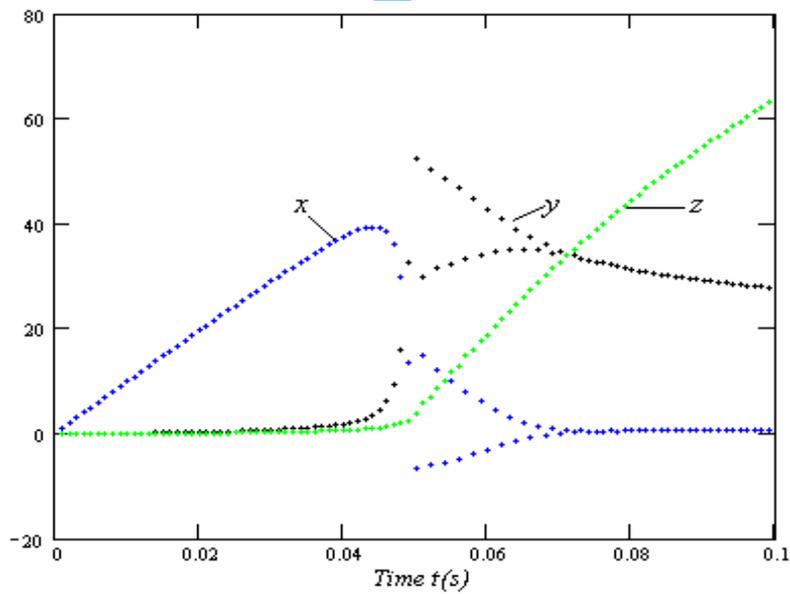

***Figure 1.3***  *is described bifurcations x, y, z components*

57

## 2 Only simple transformation of nonlinear to linear system

The coefficients leading system to a good representation of linear systems

$$c_1 = 1 \; ; \; c_2 = 1 \; ; \; c_3 = 1 \; ; \; c_4 = 1$$

$$\frac{d}{dt}x = -x - x \cdot y^2 + z + 1$$

$$\frac{d}{dt}y = x + x \cdot y^2 - y$$

$$\frac{d}{dt}z = y - z - x \cdot y \cdot z$$

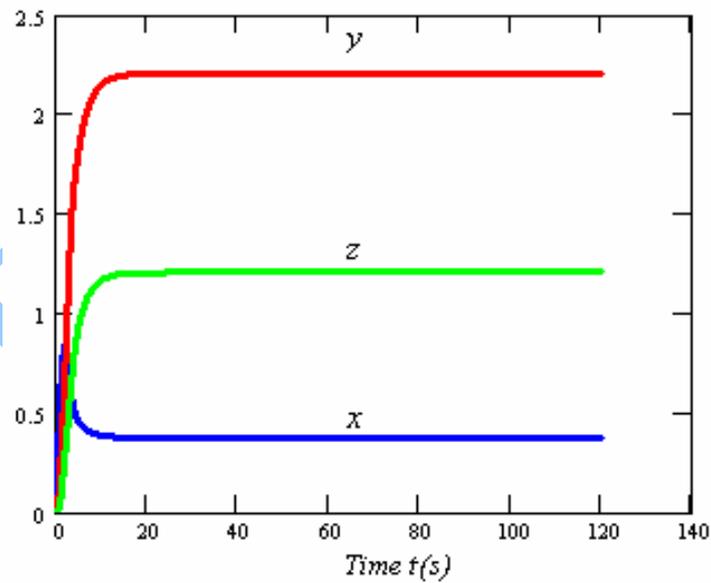

***Figure 2.1*** *get the graphics of the output of the system for a long time*

We imagine very good filters for the signals. We have for the first time a block diagram

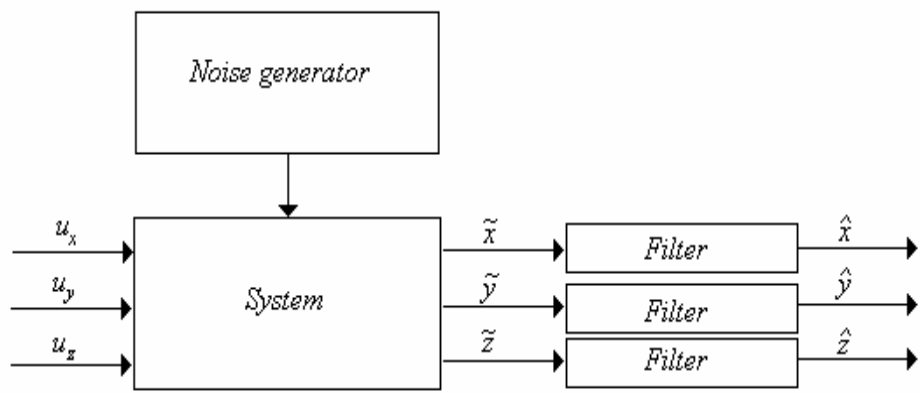

***Figure 2.2*** *the block diagram of automatic nonlinear to linear system with noise and filters*

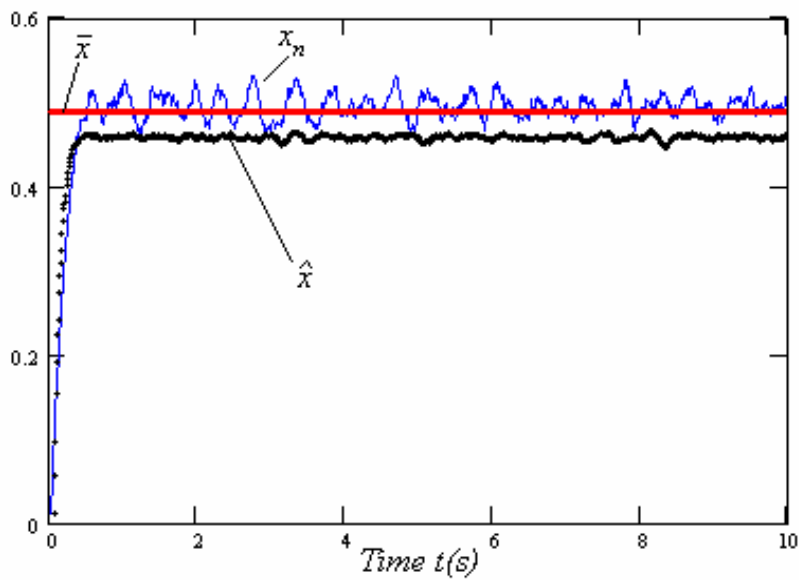

***Figure 2.3*** *one output x with noise and filter*

59

These filters are very fine controlled or regulation so, with another adjustment will be the following outputs

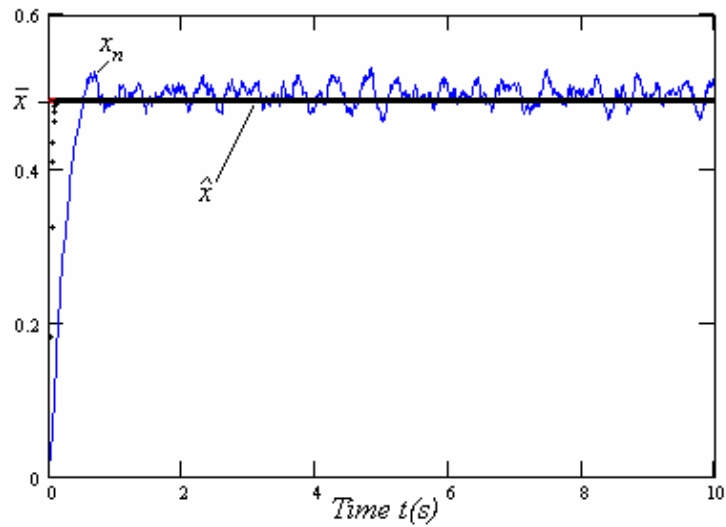

***Figure 2.4*** *one output x with noise and filter with fine adjustment*

## Conclusions

Normally, on the side we have chaos and two transforms of physical system: a new system or destroyed it means nothing.

With this sample, is important to know to find areas or pieces or ensembles of complexes in their diagnosis. For prevent so accidents or disasters.

On the other hand we obtained a linear system with other coefficients.

Was used Euler second method for numerical integration? Yes, we did!

## References

[BS98]    **V. Branzanescu and O. Stanasila**, *Special Mathematics*, All Book Company, Bucharest, 1998